# Anderson cross-localization

S. Stützer,[1] Y. V. Kartashov,[2,3,*] V. A. Vysloukh,[2] A. Tünnermann,[1] S. Nolte,[1] M. Lewenstein,[2,4]
L. Torner,[2] and A. Szameit[1]

[1]*Institute of Applied Physics, Friedrich-Schiller-Universität Jena, Max-Wien-Platz 1, 07743 Jena, Germany*
[2]*ICFO-Institut de Ciencies Fotoniques, and Universitat Politecnica de Catalunya, Mediterranean Technology Park, 08860 Castelldefels (Barcelona), Spain*
[3]*Institute of Spectroscopy, Russian Academy of Sciences, Troitsk, Moscow Region, 142190, Russia*
[4]*ICREA–Institucio Catalana de Recerca i Estudis Avançats, Lluis Companys 23, 08010 Barcelona, Spain*

We report Anderson localization in two-dimensional optical waveguide arrays with disorder in waveguide separation introduced along one axis of the array, in an uncorrelated fashion for each waveguide row. We show that the anisotropic nature of such disorder induces a strong localization along both array axes. The degree of localization in the cross-axis remains weaker than that in the direction in which disorder is introduced. This effect is illustrated both theoretically and experimentally.

Anderson localization was predicted in solid state physics upon consideration of evolution of particles in a disordered infinite medium [1]. Anderson localization appears due to transformation of infinitely extended eigenmodes of the system (Bloch waves) into exponentially localized modes in the presence of disorder [2], and it is a universal concept applicable to a variety of physical systems [3], including optics [4-7]. A breakthrough was the optical analogy of a solid and a photonic waveguide array (or lattice) where a longitudinally invariant disorder can be realized, as required for true Anderson localization [8,9]. Notice that localization is usually destroyed in the presence of longitudinal variations of disorder if their scale is smaller than the coupling length in the array [10] and may persist for large-scale variations. The effect of nonlinearity and interfaces on Anderson localization in such arrays was analyzed too [11,12].

When random fluctuations of the spacing between neighboring waveguides are introduced, the disorder is mostly found in the off-diagonals of the Hamiltonian and is therefore called *off-diagonal* disorder (ODD). If the ODD is introduced in one direction only – i.e., it is effectively one-dimensional (1D) – the eigenfunctions are mostly localized in that direction, and are still extended along the transverse direction. However, if the ODD is introduced in an uncorrelated way for each waveguide row, the effective disorder becomes anisotropically two-dimensional (2D), and therefore it can induce cross-localization along the two axes.

In this Letter we demonstrate, both theoretically and experimentally that such is the case. Namely, Anderson cross-localization (ACL) in two-dimensional setting where nominally 1D, but 2D-uncorrelated ODD is shown to be sufficient to localize the wavefunction in both transverse directions. The degree of localization in the cross-direction may be strongly enhanced and builds up fast with increasing disorder levels. Note that the quantitative aspects of this effect are quite surprising: naively, one would expect a much shorter localization length in the "strongly" disordered direction.

In order to describe ACL in disordered two-dimensional waveguide arrays we employ the nonlinear Schrödinger equation for the dimensionless light amplitude $q$:

$$i\frac{\partial q}{\partial \xi} = -\frac{1}{2}\left(\frac{\partial^2 q}{\partial \eta^2} + \frac{\partial^2 q}{\partial \zeta^2}\right) - q|q|^2 - pR(\eta,\zeta)q, \qquad (1)$$

where $\eta, \zeta$ and $\xi$ are the normalized transverse and longitudinal coordinates, respectively, the parameter $p$ describes the refractive index contrast. The refractive index distribution in disordered array is described by the function $R(\eta,\zeta) = \sum_{k,m} \exp[-(\eta-\eta_k)^4/w_\eta^4 - (\zeta-\zeta_m)^4/w_\zeta^4]$, with $w_\eta, w_\zeta$ being the widths of waveguides along the horizontal $\eta$ and vertical $\zeta$ axes, respectively. The disorder is introduced only to the $\eta$ positions of the waveguide centers and it is uncorrelated for each waveguide row along the $\zeta$ axis. Thus, while $\zeta_m = md_\zeta$ where $d_\zeta$ is the regular waveguide spacing along $\zeta$ axis and $m \in \mathbb{Z}$, the coordinates of the waveguides along the $\eta$ axis are given by $\eta_k = kd_\eta + r_{km}$, where $d_\eta$ is the regular spacing, $k \in \mathbb{Z}$, and $r_{km} < d_\eta/2$ is a random shift of the $k$-th waveguide center that is uniformly distributed in $[-S_\eta, +S_\eta]$ and that changes for different rows. The degree of disorder in such a system is controlled by the maximal possible shift of the waveguides $S_\eta$. This scheme introduces anisotropic disorder that however is much stronger along the $\eta$ axis.

In accordance with experiment we set $w_\eta = 0.3$, $w_\zeta = 0.9$ ($3 \times 9$ $\mu m^2$ wide waveguides); $d_\eta = 1.7$, $d_\zeta = 2.3$ (the separation of 17 $\mu m$ and 23 $\mu m$ along $\eta$ and $\zeta$ axes, respectively); $p = 11.3$ (the refractive index contrast $\sim 8 \times 10^{-4}$ for $\lambda = 633$ nm); the normalized total array length is $L = 34.8$ that corresponds to 50 mm of propagation. For example the disorder $S_\eta = 0.1$ allows uniformly distributed fluctuations of waveguide center along the $\eta$ axis within the segment $[kd_\eta - 1 \mu m, kd_\eta + 1 \mu m]$.

Two different approaches for the illustration of Anderson localization were adopted in the numerical simulations with Eq. (1). Using first a Monte-Carlo approach we constructed $n = 10^3$ realizations of disordered arrays for each disorder level $S_\eta$. The input beam $q|_{\xi=0} = Aw(\eta,\zeta)$, where $A$ is the input amplitude and $w(\eta,\zeta)$ is the linear mode of an isolated waveguide, was launched into the central waveguide of each array and propagated for $\xi = L$ that allows us to construct the output intensity distributions $I_{av}(\eta,\zeta) = n^{-1}\sum_{i=1,n}|q_i(\eta,\zeta,\xi=L)|^2$ averaged over the ensemble of arrays. In the frames of second approach only one extended array was constructed for each disorder

level $S_\eta$, but the averaging of output intensity distributions was performed using an excitation of different waveguides. Averaging over different waveguides (that is used in experiment) provides absolutely the same result as averaging over the ensemble of array realizations (Fig. 1), as it was recently confirmed in 1D arrays [13].

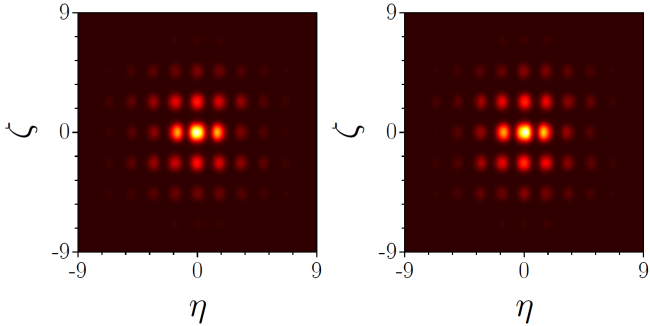

Fig. 1. (Color online) Comparison of output intensity distribution obtained by averaging over different realizations of arrays (left) and for excitation of different waveguides (right) at $S_\eta = 0.4$.

The output intensity distributions obtained by averaging over the ensemble of $n = 10^3$ arrays are shown in Fig. 2 (right column) for linear case, when $A \to 0$. At $S_\eta = 0.0$ (first row) one observes regular discrete diffraction. For small disorder levels $S_\eta \sim 0.2$ (second row) the output pattern shows a tendency for contraction: it is more confined in the horizontal direction, in which disorder is acting, and bright spots appear in the center, i.e., in this regime $I_{av}$ represents a superposition of a localized central part and several side lobes stemming from discrete diffraction. The transition to localization is apparent already at moderate disorder levels $S_\eta \sim 0.4$ (third row), when the diffraction side lobes almost vanish and only the localized central part featuring exponential tails remains. The localization is further enhanced with an increase of the disorder level up to $S_\eta \sim 0.6$ (fourth row).

The central result of our simulations is that increasing the disorder level causes simultaneous and almost equally strong Anderson localization along both $\eta$ and $\zeta$ axes (Fig. 2), despite the fact that disorder is effectively 1D (only $\eta$ coordinates of waveguides fluctuate). The localization along the $\zeta$ axis builds up with $S_\eta$ much faster than expected considering that only small disorder generated by the disorder anisotropy acts in this direction. Such a "cross-localization" effect arises because introducing a 1D disorder in the separation between sites in any multidimensional system modifies the site-to-site separation also in all orthogonal directions, affecting the coupling in these directions. The almost equal degree of localization in $\eta$ and $\zeta$ directions is particularly surprising taking into account that for $S_\eta \ll 1$ the variation of the horizontal waveguide separation $\sim S_\eta$ causes much smaller variation of the distance between waveguides in the neighboring rows $\sim S_\eta^2 / 2d_\zeta$, which intuitively should result in much weaker localization along the $\zeta$ axis. Nevertheless, a comparable localization in $\eta$ and $\zeta$ directions was obtained already at $S_\eta \sim 0.3$. The same effect is observed in arrays with circular waveguides, i.e. the enhanced localization in $\zeta$ direction is not due to elliptical waveguides.

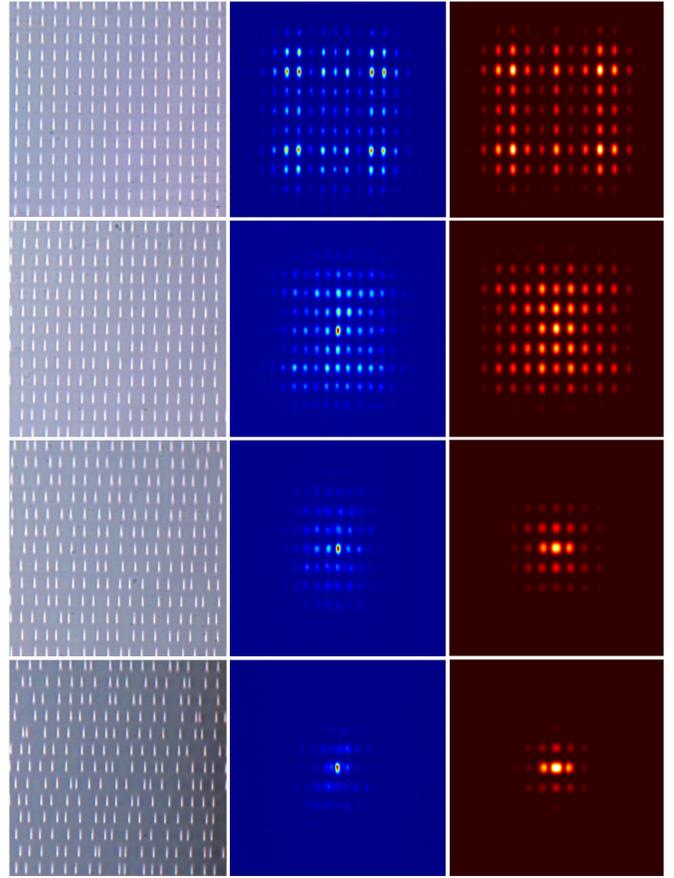

Fig. 2. (Color online) Microscopic images of disordered waveguide arrays (left), experimental (center) and theoretical (right) averaged output intensity distributions in linear case. The disorder level is $S_\eta = 0.0$, $0.2$, $0.4$ and $0.6$ from first to fourth row.

This behavior was proven in our experiments. We fabricated arrays with 21×21 single-mode waveguides for several disorder levels $S_\eta$ using the laser direct-writing technology (for the fabrication parameters see, e.g., [12]). A central section from the microscope images of waveguide arrays are shown in Fig. 2 (left column) for different disorder levels. The fluctuations in waveguide positions along the $\eta$ axis increase with growth of $S_\eta$, whereas spacing along $\zeta$ axis remains constant. Due to the fabrication procedure the waveguides are elliptical with $w_\eta < w_\zeta$, but the distances between their centers $d_\eta < d_\zeta$ were also adjusted in such a way that in regular arrays the coupling constants in $\eta$ and $\zeta$ directions are approximately equal.

To measure the output intensity distributions the linearly polarized beam at the wavelength $\lambda = 633$ nm was launched into the selected waveguide using a fiber laser and projected on a camera with $4\times$ objective. The averaged output intensity distributions are obtained by exciting well-separated waveguides in the disordered arrays with different disorder levels $S_\eta$. To collect a reliable statistics we excite 25 different waveguides, and upon averaging the centers of corresponding output distributions are adjusted in accordance with the position of excited waveguides. In all cases the excited waveguides were selected sufficiently far from the boundaries of arrays to avoid surface effects [12]. The experimental patterns shown in Fig. 2 (middle column) demonstrate a transition from discrete diffraction to strongly localized distributions with increase of disorder level. Notice a comparable degree of localization in the vertical and horizontal directions. Whereas averaged intensity distributions are localized and feature exponential tails, the individual output distributions can be weakly or strongly localized (Fig. 3) depending on which eigenmodes are excited in each realization. Subsequent beam evolution is dictated by beatings between excited eigenmodes.

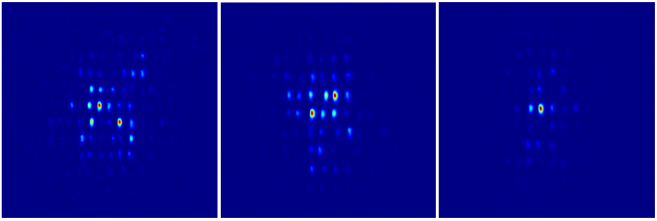

Fig. 3. (Color online) Experimental output intensity distributions for excitation of different waveguides of disordered array with $S_\eta = 0.4$ resulting in the formation of weakly (left and center) and strongly localized (right) patterns.

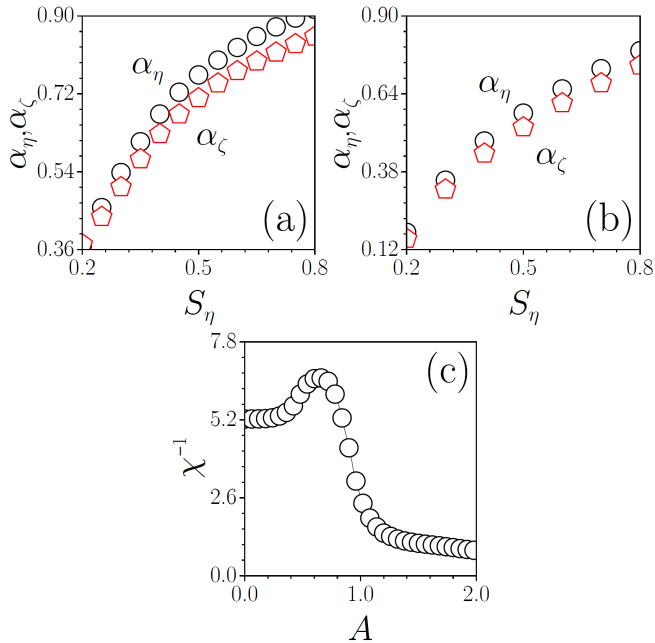

Fig. 4. (Color online) Theoretically calculated (a) and experimentally determined (b) exponential decay rates $\alpha_\eta, \alpha_\zeta$ versus disorder strength $S_\eta$. (c) Inverse form-factor of averaged output intensity distribution versus input amplitude at $S_\eta = 0.4$.

In order to compare the degrees of localization of the averaged output intensity distributions along $\eta$ and $\zeta$ axes we calculated the slopes $\alpha_\eta, \alpha_\zeta$ (or exponential decay rates) of the inner linear parts of the $\ln[I_{av}(\eta,\zeta=0)]$ and $\ln[I_{av}(\eta=0,\zeta)]$ dependencies. The theoretical and experimental exponential decay rates are shown as functions of $S_\eta$ in Figs. 4(a) and 4(b), respectively. In both experiment and theory we calculated $\alpha_\eta, \alpha_\zeta$ values starting from $S_\eta \approx 0.2$ (this is the minimum disorder level for which the appearance of an exponentially decaying inner part is ensured). The decay rates $\alpha_\eta, \alpha_\zeta$ are comparable for any disorder level, but the localization along $\eta$-direction - in which the disorder is acting - is always slightly better. Both $\alpha_\eta$ and $\alpha_\zeta$ increase monotonically with $S_\eta$.

Finally, we study the impact of nonlinearity on ACL. To this end, we calculate the inverse value of the averaged output form-factor $\chi = U^{-2} n^{-1} \sum_{i=1,n} \iint |q_i|^4 d\eta d\zeta$, where $U = \iint |q_i|^2 d\eta d\zeta$ is the total power at $S_\eta = 0.4$ for $n = 10^3$ array realizations as a function of input beam amplitude $A$. The quantity $\chi^{-1}$ provides the information about the width of the central "localized" portion of the output wavepacket and disregards small-amplitude diffracting waves. First, $\chi^{-1}$ slightly increases with increasing peak amplitude (i.e., initially focusing nonlinearity can even provide a small delocalization), but when $A$ reaches a certain threshold level the width of localized region rapidly decreases indicating that on the length of our sample the focusing nonlinearity stimulates localization [Fig. 4(c)]. For high peak amplitudes $\chi^{-1}$ slowly decreases, but in this case almost all light is localized in the excited guide and disorder is too weak to compete with nonlinearity.

Summarizing, we demonstrated that uncorrelated off-diagonal 1D disorder in 2D waveguide array results in Anderson cross-localization in both transverse directions. Such anisotropic off-diagonal disorder couples both transverse dimensions so that the localization along the direction in which the effective disorder is only very weak builds up remarkably fast.